\documentclass[11pt]{article}
\usepackage{moriond,epsfig}

\def\be{\begin{equation}}
\def\ee{\end{equation}}
\def\bea{\begin{eqnarray}}
\def\eea{\end{eqnarray}}

\begin{document}
\vspace*{4cm}
\title{ON THE OBSERVATIONAL STATUS OF ULTRAHIGH ENERGY COSMIC RAYS
AND THEIR POSSIBLE ORIGIN IN STARBURST-LIKE GALAXIES}

\author{Diego F. Torres$^a$ and Luis A. Anchordoqui$^b$}

\address{$^a$Lawrence Livermore National Laboratory, 7000 East Ave., L-413,
Livermore, CA 94550, USA\\
$^b$ Department of Physics, Northeastern University,
Boston MA 02115, USA}

\maketitle\abstracts{ This paper presents a brief review of the
current status of ultrahigh energy cosmic ray
observations and discusses nearby starburst-like galaxies as their
possible origin. }

\section{Energy Spectrum, Mass Composition, and Distribution of Arrival Directions}

In the almost structureless, fast falling with energy, inclusive
cosmic ray (CR) spectrum, three kinematic features have drawn
considerable attention for a long time. These features, known as
the knee, the ankle, and the ultraviolet cutoff, are the only ones
in which the spectral index shows a sharper variation as a
function of energy, probably signaling some ``new physics''. A
plethora of astrophysical~\cite{Torres:2004hk} and
exotic~\cite{Anchordoqui:2002hs} explanations have been proposed
to address the production mechanism at the high end of the
spectrum. In the absence of a single model which is consistent
with all data, the origin of these particles remains a mystery.
Clues to solve it are not immediately forthcoming from the data,
particularly since various experiments report mutually
inconsistent results.

In recent years, a somewhat confused picture of the energy
spectrum has been emerging. Since 1998, the AGASA Collaboration
has consistently reported~\cite{Takeda:1998ps} a continuation of
the spectrum beyond the expected Greisen--Zatsepin--Kuzmin (GZK)
cutoff,\cite{Greisen:1966jv} which should arise at about
$10^{10.7}$~GeV if CR sources are at cosmological
distances. In contrast, the most recent results from
HiRes~\cite{Abbasi:2002ta} describe a spectrum which is consistent
with the expected GZK feature. This situation exposes the
challenge posed by systematic errors (predominantly arising from
uncertainties in hadronic interaction
models~\cite{Anchordoqui:1998nq}) in these types of measurements.

Recent HiRes data have been interpreted as a change in CR
composition, from heavy nuclei to protons, at $\sim
10^9$~GeV.\cite{Bergman:2004bk} This is an order of magnitude
lower in energy than the previous crossover deduced from the Fly's
Eye data.\cite{Bird:1993yi} The end-point of the galactic flux is
expected to be dominated by iron, as the large charge $Ze$ of
heavy nuclei reduces their Larmor radius (containment scales
linearly with $Z$) and facilitates their acceleration to highest
energy (again scaling linearly with $Z$). The dominance of nuclei
in the high energy region of the Galactic flux carries the
implication that any changeover to protons represents the onset of
dominance by an extra-galactic component. The  inference from this
new HiRes data is therefore that the extra-galactic flux is
beginning to dominate the Galactic flux already at $\sim
10^9$~GeV. Significantly, this is well below $E_{\rm GZK} \sim
10^{10.7}$~GeV, the threshold energy for resonant $p \gamma_{\rm
CMB} \rightarrow \Delta^+ \rightarrow N \pi$ energy-loss on the
cosmic microwave background (CMB), and so samples sources even at
large redshift.

The dominance of extra-galactic protons at lower energy can be
consistent with recently corroborated structures in the CR
spectrum.  A second knee, recognized originally in AGASA
data~\cite{Nagano:1991jz} is now confirmed by the HiRes-MIA
Collaboration.\cite{Abu-Zayyad:2000ay} At $10^{8.6}$~GeV, the
energy spectrum steepens from $E^{-3}$ to $E^{-3.3}$. This
steepening at the second knee can be
explained~\cite{Berezinsky:2005cq} by energy losses of
extra-galactic protons over cosmic distances, due to $e^+ e^-$
pair-production on the CMB. The theoretical threshold of the
energy-loss feature occurs at $10^{8.6}$~GeV, and therefore allows
for proton dominance even below this energy. However, the HiRes

coincident with the energy of the second knee (from about 50\%
protons just below to 80\% protons just above), and therefore
argues for the beginning of extra-galactic proton dominance at the
second knee. Another feature in the CR  spectrum is the
ankle at $\sim 10^{10}$~GeV where the spectrum flattens from
$E^{-3.3}$ to $E^{-2.7}$. This has been commonly identified with
the onset of the extra-galactic flux in the past. In the aftermath
of the new HiRes data, the ankle can now be interpreted as the
minimum in the $e^+ e^-$ energy-loss feature. These changes in
viewing the onset of the extra-galactic component have spurred a
refitting of the CR data down to $10^{8.6}$~GeV with
appropriate propagation functions and extra-galactic injection
spectra.\cite{Ahlers:2005sn} The major result is that the
injection spectrum is significantly steeper than the standard
$\propto E^{-2}$ predicted by Fermi engines. This appears to
contradict the latest HESS measurements of the spectral
slope of both, identified and unidentified gamma-ray sources (albeit at
lower energies, see M. Lemoine-Gourmard in this proceedings).

At the highest energies, the arrival directions of CRs are
expected to begin to reveal their origins. If the CR
intensity were isotropic, then one should expect a
time-independent flux from each direction in local detector
coordinates, i.e., declination and hour angle.  In that case, a
shower detected with local coordinates could have arrived with
equal probability at any other time of a shower detection. For any
point of the celestial sphere, the expected shower density can be
estimated if the exposure in each direction can be obtained.  This
implies that celestial anisotropies can be easily discerned by
comparing the observed and expected event frequencies at each
region. Although there seems to be a remarkable agreement among
experiments on predictions about isotropy on large scale
structure,\cite{Anchordoqui:2003bx} this is certainly not the case
when considering the two-point correlation function on small
angular scale. The AGASA Collaboration reports observations of
event clusters which have a chance probability smaller than 1\% to
arise from a random distribution,\cite{Hayashida:bc} whereas the
recent analysis reported by the HiRes Collaboration showed that
their data are consistent with no clustering among the highest
energy events.\cite{Abbasi:2004ib} In our opinion, it is {\it very
important} to rigorously define the corresponding budget of
statistical significance and search criteria {\em prior} to
studying the data, since defining them {\em a posteriori} may
inadvertently introduce an undetermined number of ``trials'' and
thus make it impossible to assign the correct statistical
significance to the search result. In this direction,  with the
aim of avoiding accidental bias on the number of trials performed
in selecting the angular bin, the original claim of the AGASA
Collaboration was re-examined considering only those events
observed after the original claim.\cite{Finley:2003ur} This study
showed that the evidence for clustering in the AGASA data set is
weaker than was previously supposed, and is consistent with the
hypothesis of isotropically distributed arrival directions.

On a separate track, evidence has  been presented for neutral
particles (with energy $\sim 10^{9}$~GeV) from the Cygnus spiral
arm. Specifically, AGASA has revealed a correlation of the arrival
direction of the CRs to the Galactic Plane (GP) at the
$4\sigma$ level.\cite{Hayashida:1998qb}  The GP excess, which is
roughly 4\% of the diffuse flux, is mostly concentrated in the
direction of the Cygnus region.\cite{Teshimaicrc} Evidence at the
3.2$\sigma$ level for GP enhancement in a similar energy range has
also been reported by the Fly's Eye
Collaboration.\cite{Bird:1998nu}  Interestingly, the full Fly's
Eye data include a directional signal from the Cygnus region which
was somewhat lost in an unsuccessful attempt to relate it to
$\gamma$--ray emission from Cygnus X-3.\cite{Cassiday:kw} The
complete isotropy up to about $10^{7.7}$~GeV revealed by KASCADE
data~\cite{Antoni:2003jm} vitiate direction-preserving photons as
primaries. Therefore, the excess from the GP is very suggestive of
neutrons as candidate primaries, because the directional signal
requires relatively-stable neutral primaries, and time-dilated
neutrons can reach the Earth from typical Galactic distances when
the neutron energy exceeds $10^{9}$~GeV. If the Galactic
messengers are neutrons, then those with energies below
$10^{9}$~GeV will decay in flight, providing a flux of cosmic
antineutrinos above 1~TeV that should be observable at
kilometer-scale neutrino telescopes.\cite{Anchordoqui:2003vc} A
measurement of the $\bar \nu$-flux will supply a strong
confirmation of the GP neutron hypothesis.

Another point of agreement among the experiments is the absence of a photon component at the
highest energies.\cite{Ave:2000nd} This was most recently supported by an analysis of AGASA data: above
$10^{11.2}$~GeV, less than 50\% (65\%)
of the primary CRs can be photons at 90\%CL (95\% CL).\cite{Risse:2005jr}

\begin{figure}
%\rule{5cm}{0.2mm}\hfill\rule{5cm}{0.2mm}
%\vskip 2.5cm
%\rule{5cm}{0.2mm}\hfill\rule{5cm}{0.2mm}
\begin{center}
\psfig{figure=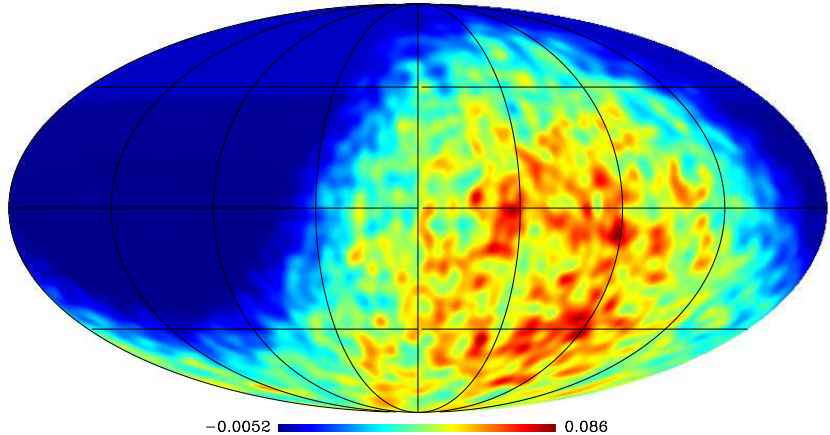,height=2.8in}
\end{center}
\caption[...]{All event (from January to July 2004) skymap in Galactic
coordinates. Units $x$ are related to the
number of events per pixel $n$ according to: $n = 330 \,x + 1.716$.\cite{Anchordoqui:2004wb}
\label{skymap}}
\end{figure}

The confusing experimental situation regarding the GZK
feature, mass composition, and the small-scale clustering in the distribution of arrival directions
should be resolved by the Pierre Auger Observatory (PAO),
which will provide not only a data set of
unprecedented size, but also the machinery for controlling some of the more problematic systematic
uncertainties.
The first PAO site is now operational in Malarg\"ue, Argentina, and is in the process
of growing to its final size of 3000~km$^2$.\cite{Abraham:2004dt}
At the time of writing, 12 telescopes and about 700 water tanks were operational.
The first analyses of data from the PAO are currently underway.\cite{Anchordoqui:2004wb}
 Figure~\ref{skymap} shows the
arrival directions of all events recorded from January to July 2004. The pixels have a size of 1.8
degrees and the map was smoothed with a Gaussian beam of 5 degrees. On 21 May 2004,
one of the larger events recorded by the surface array triggered 34 stations. A preliminary estimate
yields an energy~$\sim~10^{11}$~GeV and a zenith angle of about $60^\circ.$ First physics results
will be made public in the Summer of 2005 at the 29th International Cosmic Ray Conference.

\section{Possible Astrophysical Origin}

Supernova remnants (SNRs) are thought to be the main source of
both CR ions and electrons with energies below the knee (see Ref.\cite{Torres:2002af}
for a review, and references therein for
details). CRs of low energies are also expected to be accelerated
in OB associations, through turbulent motions and collective
effects of star winds. When the Larmor radius $r_{\rm L}$
approaches the accelerator size, it becomes very difficult to
magnetically confine the CR to the acceleration region, and thus
to continue the accelerating process up to higher energies. If one
includes the effect of the characteristic velocity $\beta c$ of
the magnetic scattering centers, the above argument leads to the
general condition, $ E_{\rm max} \sim 2 \beta\, c\, Ze\,B\,
r_{_{\rm L}}\,,$ for the maximum energy acquired by a particle
travelling in a medium with magnetic field $B$.
In the case of one-shot acceleration scenarios, the maximum
reachable energy turns out to have a quite similar expression to
the shock acceleration case.
In what follows, we focus on only one possible astrophysical
origin of ultrahigh energy CRs, and refer the reader to~\cite{Torres:2004hk} for
thorough discussion on other candidates.

\subsection{The two nearest starbursts}

Starbursts are galaxies  undergoing a large-scale star formation
episode. They feature  strong infrared emission originating in the
high levels of interstellar extinction, strong HII-region-type
emission-line spectrum (due to a large number of O and B-type
stars), and  considerable radio emission produced by recent SNRs.
Typically, starburst regions are located close to the galactic
center, in the central kiloparsec. This region alone can be orders
of magnitude brighter than the center of normal spiral galaxies.
From such an active region, a galactic-scale superwind is driven
by the collective effect of supernovae and particular massive star
winds. The enhanced supernova explosion rate creates a cavity of
hot gas ($\sim10^8$ K) whose cooling time is much greater than the
expansion time scale. Since the wind is sufficiently powerful, it
can blow out the interstellar medium of the galaxy, preventing it
from remaining trapped as a hot bubble. As the cavity expands, a
strong shock front is formed on the contact surface with the cool
interstellar medium. The shock velocity can reach several
thousands of kilometers per second and ions like iron nuclei can
be efficiently accelerated in this scenario, up to ultrahigh
energies, by Fermi's mechanism.\cite{Anchordoqui:1999cu} If the
super-GZK particles are heavy nuclei from outside our Galaxy, then
the nearby ($\sim 3$ Mpc~\cite{heckman}) starburst galaxies M82
($l=141^\circ, b=41^\circ$) and NGC 253 ($l=89^\circ, b=
-88^\circ$) are prime candidates for their origin.

M82 is probably the best studied starburst galaxy, located at 3.2
Mpc. The total star formation rate in the central parts is at
least $\sim10$ M$_{\odot}$ yr$^{-1}.$\cite{OM} The far infrared
luminosity of the inner region within 300 pc of the nucleus is
$\sim 4\times 10^{10}$ L$_{\odot}$.\cite{rieke} There are $\sim
1\times 10^7$ M$_{\odot}$ of ionized gas and $\sim 2 \times 10^8$
M$_{\odot}$ of neutral gas in the IR source.\cite{rieke,SA} The
total dynamical mass in this region is $\sim (1-2) \times 10^9$
M$_{\odot}$.\cite{SA} The main observational features of the
starburst can be modelled with a Salpeter IMF extending from 0.1
to 100 M$_{\odot}$. The age of the starburst is estimated in $\sim
(1-3) \times 10^7$ yr.\cite{rieke} Around $\sim 2.5\times 10^8$
M$_{\odot}$ (i.e. $\sim 36$ \% of the dynamical mass) is in the
form of new stars in the burst.\cite{SA} The central region,
then, can be packed with large numbers of early-type stars.
More than 60 individual compact radio sources have been detected
within the central 200 pc of NGC 253,\cite{ulvestad} half of
which are SNRs. The supernova rate is
estimated to be as high as $0.2-0.3$ yr$^{-1}$, comparable to the
massive star formation rate, $\sim 0.1$M$_\odot$ yr$^{-1}$.
Assuming that the star formation rate has been continuous in the
central region for the last 10$^9$ yrs, and a Salpeter IMF for
0.08-100 $M_\odot$, the bolometric luminosity of NGC 253 is
consistent with 1.5 $\times 10^8 M_\odot$ of young stars.\cite{watson}
Based on this evidence, it appears likely that
there are at least tens of millions of young stars in the central
region of the starburst. These stars can  contribute to the
$\gamma$-ray luminosity at high
energies.\cite{Romero:2003tj,Torres:2003ur} A TeV detection was
reported by CANGAROO~\cite{Itoh:2003yg}, but has been yet
unconfirmed by other experiments. A multiwavelength model of NGC
253, predicting less gamma-ray flux than that detected by CANGAROO
(thus, predicting that HESS will not detect the starburst) has
been presented in Ref.\cite{NGC253}

\subsection{Acceleration processes in starbursts}

 Due to the
nature of the central region, and the presence of the superwind,
the escape of the iron nuclei from the central region of the
galaxy is dominated by convection.\footnote{The relative
importance of convection and diffusion in the escape of the CRs
from a region of disk scale height $h$ is given by the
dimensionless parameter, $q={V_0\,h}/{\kappa_0}, $ where $V_0$ is
the convection velocity and $\kappa_0$ is the CR diffusion
coefficient inside the starburst.\cite{new3} When $q < 1$, the CR
outflow is difussion dominated, whereas when $q > 1$ it is
convection dominated. For the central region of NGC 253 a
convection velocity of the order of the expanding SNR shells
$\sim$ 10000 km s$^{-1}$, a scale height $h \sim 35$ pc, and a
reasonable value for the diffusion coefficient $\kappa_0 \sim 5
\times 10^{26}$ cm$^2$ s$^{-1}$, lead to $q
\sim 216$. Thus, convection dominates the escape of the particles.
The residence time of the iron nuclei in the starburst results
$t_{\rm RES} \sim h / V_0 \approx 1 \times 10^{11}$ s.}  Nuclei
can then escape through the disk in opposite directions along the
symmetry axis of the system, being the total path travelled
substantially shorter than the mean free path. Once the nuclei
escape from the central region of the galaxy  they are injected
into the galactic-scale wind and experience further acceleration
at its terminal shock. CR acceleration at superwind shocks was
first proposed in~\cite{bebito} in the context of our own Galaxy.
The scale length of this second shock is of the order of several
tens of kpc,\cite{heckman} so it can be considered as locally
planar for calculations. The shock velocity $v_{\rm sh}$ can be
estimated from the empirically determined superwind kinetic energy
flux $\dot{E}_{\rm sw}$ and the mass flux $\dot{M}$ generated by
the starburst through: $ \dot{E}_{\rm sw}={1}/{2} \dot{M} v_{\rm
sh}^2. $ The shock radius can be approximated by $r\approx v_{\rm
sh} \tau$, where $\tau$ is the starburst age. Since the age is
about a few tens of million years, the maximum energy attainable
in this configuration is constrained by the limited acceleration
time arising  from the finite shock's lifetime. The photon field
energy density drops to values of the order of the CMB, and
consequently, iron nuclei are safe from
photodissociation while energy increases to $\sim 10^{11}$~GeV.

To estimate the maximum energy that can be reached by the nuclei,
consider the superwind terminal shock propagating in a homogeneous
medium with an average magnetic field $B$. If we work in the frame
where the shock is at rest, the upstream flow velocity will be
${\bf v_1}$ ($|{\bf v_1}|=v_{\rm sh}$) and the downstream
velocity, ${\bf v_2}$. The magnetic field turbulence is assumed to
lead to isotropization and consequent diffusion of energetic
particles which then propagate according to the standard transport
theory.\cite{jokipii} The acceleration time scale is then:
$ t_{\rm acc}=\frac{4 \kappa}{v_1^2}  \label{t} $
where $\kappa$ is the upstream diffusion coefficient which can be
written in terms of perpendicular and parallel components to the
magnetic field, and the angle $\theta$ between the (upstream)
magnetic field and the direction of the shock propagation, $
\kappa=\kappa_{\parallel} \cos^2\theta + \kappa_{\perp}
\sin^2\theta$.\cite{drury} Since strong turbulence is expected from the
shock we can take the Bohm limit for the upstream diffusion
coefficient parallel to the field, i.e. $
\kappa_{\parallel}=\frac{1}{3}{E}/{ZeB_1} ,$ where $B_1$ is the
strength of the pre-shock magnetic field and $E$ is the energy of
the $Z$-ion. For the $\kappa_{\perp}$ component we shall assume,
following Biermann,\cite{birmanncr1} that the mean free path
perpendicular to the magnetic field is independent of the energy
and has the scale of the thickness of the shocked layer ($r/3$).
Then, $ \kappa_{\perp}={1}/{3} \; r (v_1-v_2) $ or, in the strong
shock limit, $ \kappa_{\perp}={r v_1^2}/{12}.$  The upstream time
scale is $t_{\rm acc}\sim r/(3 v_1)$,
${r}/{3v_1}={4}/{v_1^2}\left({E}/({3ZeB_1}) \cos^2\theta + {r
v_1^2}/{12} \sin^2\theta\right). $ Thus, using $r=v_1\tau$ and
transforming to the observer's frame one obtains
$
E_{\rm max}\approx\frac{1}{4} ZeB v_{\rm sh}^2 \tau
\approx\frac{1}{2} ZeB \frac{\dot{E}_{\rm sw}}{\dot{M}} \tau .
$

The predicted kinetic energy and mass fluxes of the starburst of
NGC 253 derived from the measured IR luminosity are
$2\times10^{42}$ erg s$^{-1}$ and 1.2 M$_{\odot}$ yr$^{-1}$,
respectively.\cite{heckman} The age is in the range $5\times
10^7$ to $1.6\times 10^8$ yr (also valid for M82).\cite{rieke}
Finally, the radio and $\gamma$-ray emission from NGC 253 are well
matched by models with $B\sim50\mu$G,\cite{paglione} although it
might be at least a factor of 4 bigger in the innermost region of
the nucleus.\cite{NGC253} With these figures, assuming a
conservative age $\tau=50$ Myr, one obtains a maximum energy for
iron nuclei of $ E_{\rm max}^{\rm Fe}
> 3.4\times 10^{20}\;{\rm eV}.$

\subsection{Testing the starburst hypothesis}

For an extragalactic, smooth, magnetic field of $\approx 15
-20$~nG, diffusive propagation of nuclei below $10^{20}$~eV
evolves to nearly complete isotropy in the CR arrival
directions.\cite{Anchordoqui:2001ss,Bertone:2002ks} Thus, we
could use the rates at which starbursts inject mass, metals and
energy into superwinds to get an estimate of the CR-injection
spectra. Using equal power per decade over the interval
$10^{18.5}\,{\rm eV}< E< 10^{20.6}\,{\rm eV}$ -- we obtain a
source CR-luminosity
$
\frac{E^2 \,dN_0}{dE\,dt} \, \approx 3.5 \,\varepsilon \,10^{53}
{\rm eV/s} \label{1crluminosity}
$
where  $\varepsilon$ is the efficiency of ultrahigh energy CR
production by the superwind kinetic energy flux. With this in
mind, the energy-weighted, approximately isotropic nucleus flux at
$10^{19}$ eV is given by~\cite{Anchordoqui:2001ss}
$
E^3 J(E)  =  \frac{Ec}{(4\pi)^2d\,D(E)} \frac{E^2
\,dN_0}{dE\,dt}\, I_\star
  \approx  2.3 \times 10^{26} \, \epsilon \,
I_\star \, {\rm eV}^2 \, {\rm m}^{-2} \, {\rm s}^{-1} \, {\rm
sr}^{-1},
$
where $I_\star = I_{\rm M82} + I_{\rm NGC\ 253}$. To estimate the
diffusion coefficient we used $B_{\rm nG} = 15$, $\ell_{\rm Mpc} =
0.5$, and an average $Z=20$. We fix
$
\epsilon\, I_\star =0.013, \label{mimi}
$
after comparing the above equations to the observed CR-flux. Note that
the contribution of $I_{\rm M82}$ and $I_{\rm NGC\ 253}$ to
$I_\star$ critically depends on the age of the starburst. The
relation ``starburst-age/superwind-efficiency'' derived from
Eq.~(\ref{mimi}), leads to $\epsilon \approx 10\%$, if both M82
and NGC 253 were active for $115$~Myr. The power requirements may
be reduced assuming contributions from M82
``B''.\cite{Anchordoqui:2001ss}

\begin{figure}
\begin{center}
\includegraphics[width=6cm,height=7cm]{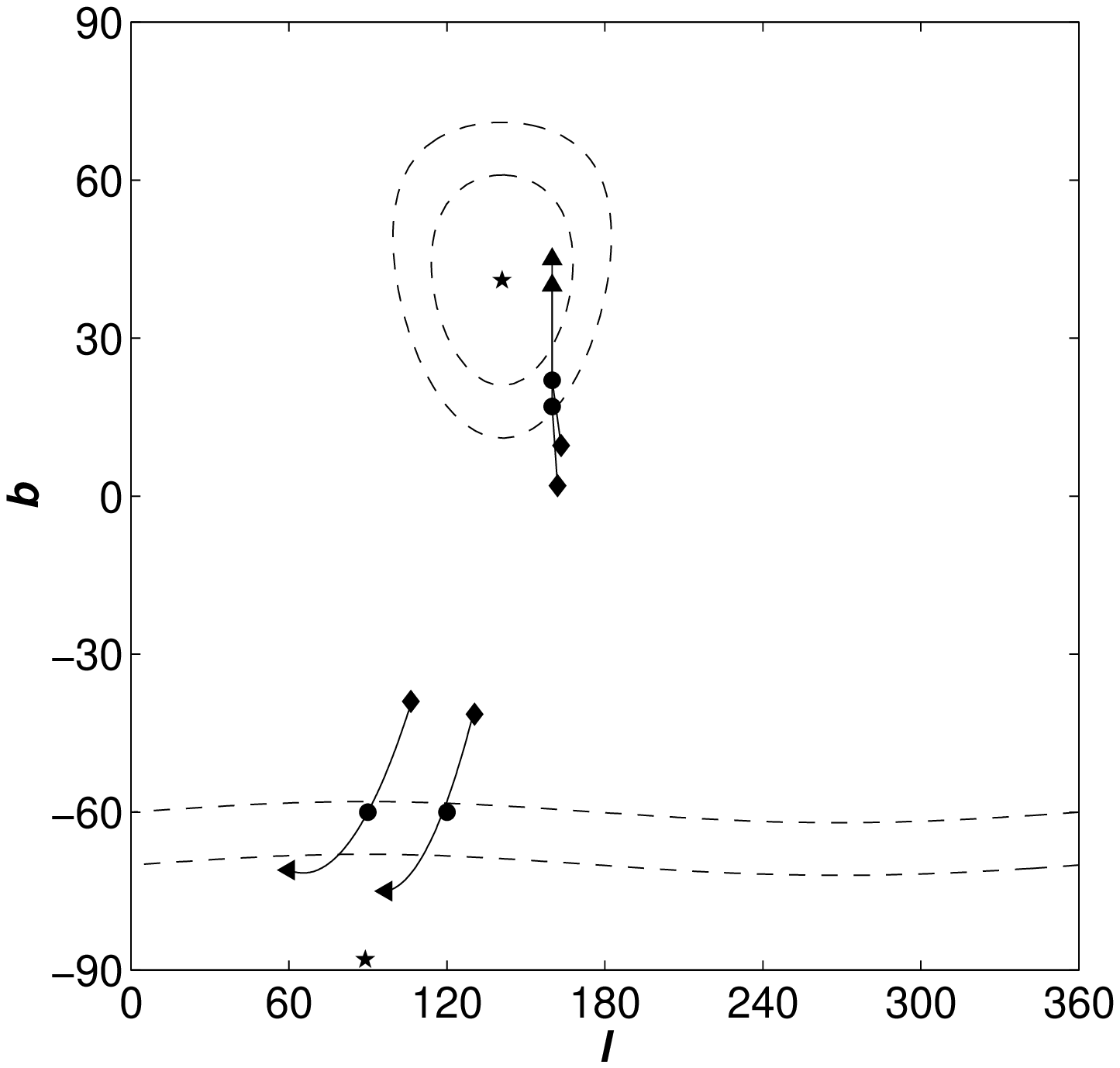}
\includegraphics[height=7cm]{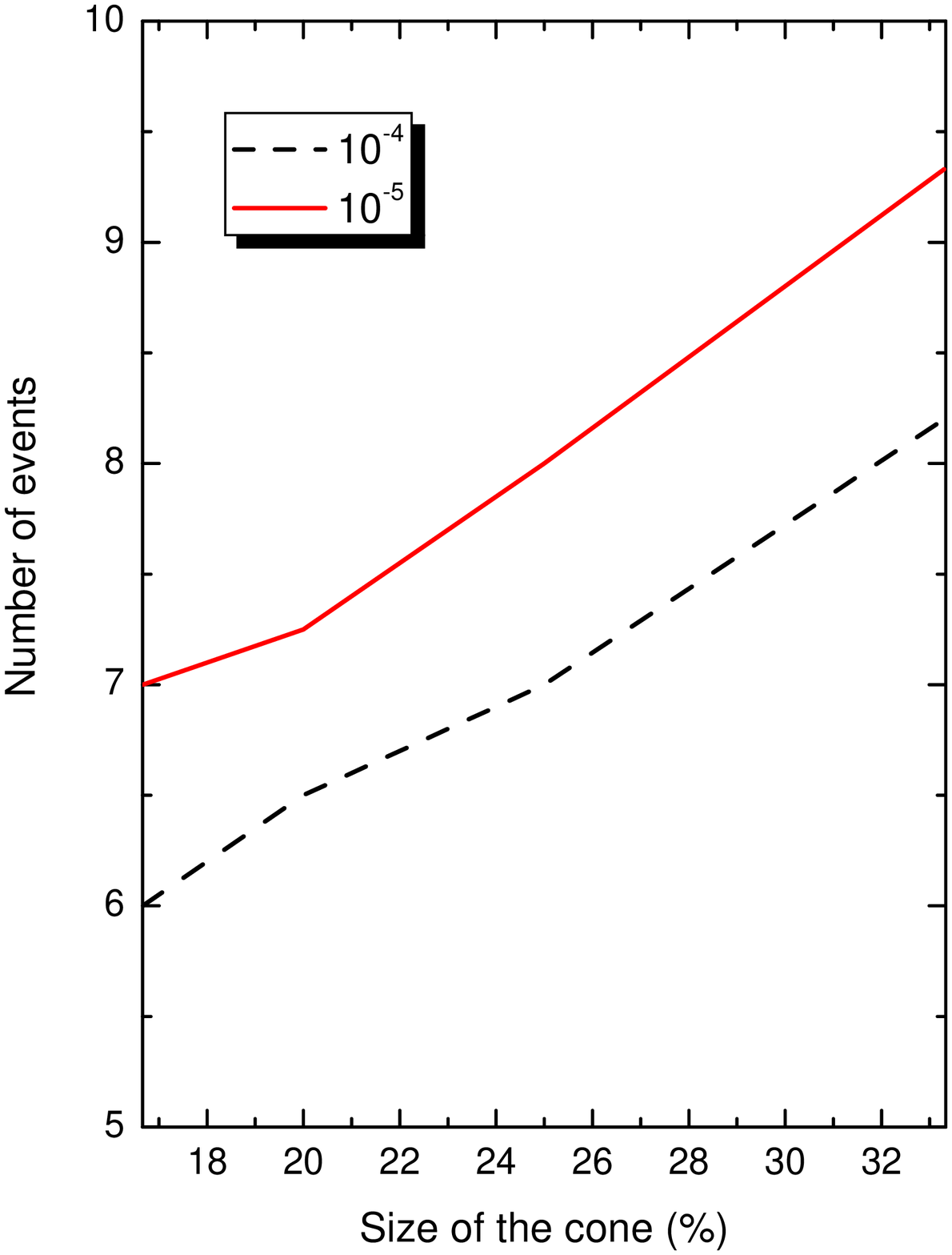}
\caption{Left: Directions in Galactic coordinates of the four
highest energy CRs at the boundary of the Galactic halo. The
diamonds represent the observed incoming directions. The circles
and arrows show the directions of neon and iron nuclei,
respectively, before deflection by the Galactic magnetic field.
The solid line is the locus of  incoming directions at the halo
for other species with intermediate atomic number. The stars
denote the positions of M82 and NGC253. The dashed lines are
projections in the $(l,b)$ coordinates of angular directions
within $20^{\circ}$ and $30^{\circ}$ of the starbursts. Right:
Curves of constant probabilities in the two-dimensional parameter
space defined by the size of the cone and the minimum number of
events originating within the resulting effective solid angle.}
\label{magnetica}
\end{center}
\end{figure}

In the non-diffusive regime (i.e., $10^{20.3}~{\rm eV} < E <
10^{20.5}~{\rm eV}$), the accumulated deflection angle from the
direction of the source in the extragalactic $B$-field is roughly
$10^\circ < \theta < 20^\circ$.\cite{Bertone:2002ks} The nuclei
suffer additional deflection in the Galactic magnetic field. In
particular, if the Galactic field is of the ASS type, the arrival
direction of the 4 highest energy CRs can be traced backwards to
one of the starbursts.\cite{Anchordoqui:2002dj}
Figure~\ref{magnetica} shows the extent to which  the observed
arrival directions of the highest energy CRs deviate from their
incoming directions at the Galactic halo because of bending in the
magnetic field. It is  seen that trajectories for CR nuclei  with
$Z\ge 10$  can be further traced back to one of the starbursts,
within the uncertainty of the extragalactic deviation.

We now attempt to assess to what extent these correlations are
consistent with chance coincidence. The deflections in the
extragalactic and Galactic fields (regular and random components)
may be assumed to add in quadrature, so that the angular sizes of
the two sources are initially taken as cones with opening
half-angles between 40$^{\circ}$ and 60$^{\circ}$, which for the
purpose of our numerical estimate we  approximate to 50$^\circ$.
The global structure of the field will substantially diminishing
the effective solid angle. The combined deflections in the $l$ and
$b$ coordinates  concentrate the effective angular size of the
source to a considerably smaller solid angle. As a conservative
estimate, we retain 25\% of this cone as the effective source
size.
By randomly generating 4 CR positions in the portion of the sky
accessible to the existing experiments (declination range $\delta
> -10^\circ$), an expected number of random coincidences can be
obtained. The term ``coincidence'' is herein used to label a
synthetic CR whose position in the sky lies within an effective
solid angle $\Omega_{\rm eff}$ of either starburst. $\Omega_{\rm
eff}$ is characterized by a cone with opening half-angle reduced
from $50^{\circ}$ to $24^{\circ}$ to account for the 75\%
reduction in effective source size due to the magnetic biasing
discussed above.  For the 4 observed events, with zero background,
the Poisson signal mean 99\% confidence interval is $0.82-12.23$.
Thus our observed mean for random events, $0.81 \pm 0.01$, falls
at the lower edge of this interval, yielding a 1\% probability for
a chance occurrence.
Assuming an extrapolation of AGASA flux ($E^3 J_{\rm obs} (E)$) up
to $10^{20.5}$~eV, the event rate at PAO, with an aperture $A
\approx 7000$~km$^2$ sr for showers with incident zenith angle
less than $60^\circ$, is given by
$
\frac{dN}{dt}  =  A\, \int_{E_1}^{E_2}\, E^3 J(E)\, \frac{dE}{E^3}
   \approx  \frac{A}{2} \,\langle E^3\, J(E) \rangle\,
  \left[ \frac{1}{E_1^2} - \frac{1}{E_2^2} \right]
 \approx  5.3 \,\,{\rm yr}^{-1}\,\,,
$ where $E_1 = 10^{20.3}$~eV and $E_2 = 10^{20.5}$~eV. Considering
a 5-year sample of 25 events and that for this energy range the
aperture of PAO is mostly receptive to CRs from NGC253, we
allow for different possibilities of the effective reduction of
the cone size because of the Galactic magnetic field biasing
previously discussed. In Fig.~\ref{magnetica} we plot contours of
constant probabilities ($P= 10^{-4},\ 10^{-5}$) in the
two-dimensional parameter space of the size of the cone (as a
fraction of the full $50^{\circ}$ circle) and the minimum number
of events originating within the resulting effective solid angle.
The model predicts that after 5 years of operation, even if 7 are
observed, it would rule out a random fluctuation at the $10^{-5}$
level.

Other galaxies, presenting a more extreme starburst behavior, like
those luminous and ultra-luminous infrared galaxies which are also
expected to be detected by space and ground-based gamma-ray
experiments,\cite{Torres:2004wf} may
additionally contribute to the ultrahigh energy CR flux. \\

\section*{Acknowledgments}

The work of DFT was performed under the auspices of the U.S.
Department of Energy (NNSA) by University of California's LLNL
under contract No. W-7405-Eng-48. He acknowledges the European
Union for the grant that allowed his participation in this
meeting, and the organizers for hospitality. The work of LAA has
been partially supported by the US National Science Foundation
(NSF) under grant No. PHY-0140407.

\end{document}